\documentclass[a4paper,11pt]{article}
\pdfoutput=1
\usepackage{jcappub}
\usepackage{aas_macros}

\usepackage{hyperref}
\hypersetup{colorlinks=true, citecolor=blue, urlcolor=blue, linkcolor=blue}

\usepackage[utf8]{inputenc}
\usepackage[T1]{fontenc}
\usepackage{graphicx}
\usepackage[normalem]{ulem}

\usepackage{cleveref,natbib,amsmath,amssymb,bm,mathrsfs}

\usepackage{siunitx}
\DeclareSIUnit{\parsec}{pc}
\DeclareSIUnit{\year}{yr}
\DeclareSIUnit{\au}{au}

\newcommand{\obj}{S}
\newcommand{\du}{\mathrm{d}}
\newcommand{\bb}[1]{\bm{\mathrm{#1}}}
\newcommand{\arctanh}{\operatorname{arctanh}}
\newcommand{\kB}{k_{\mathrm{B}}}
\newcommand{\vesc}{v_{\mathrm{esc}}}
\newcommand{\scap}{\sigma_{\mathrm{cap}}}
\newcommand{\Rej}{R_{\mathrm{ej}}}
\newcommand{\Rcap}{R_{\mathrm{cap}}}
\newcommand{\loss}{E_{\mathrm{loss}}}
\newcommand{\mach}{\mathcal{M}}
\newcommand{\pbh}{\mathrm{PBH}}
\newcommand{\target}{\mathrm{target}}

\newcommand{\Refcite}[1]{Ref.~\cite{#1}}
\newcommand{\refcite}[1]{ref.~\cite{#1}}
\newcommand{\refscite}[1]{refs.~\cite{#1}}

\title{Capture of primordial black holes in extrasolar systems}

\author[a,b]{Benjamin V. Lehmann,}
\author[a,b]{Ava Webber,}
\author[c]{Olivia G. Ross}
\author[a,b]{and Stefano Profumo}

\affiliation[a]{Department of Physics, University of California, Santa Cruz\\ 1156 High St., Santa Cruz, CA 95064, USA}
\affiliation[b]{Santa Cruz Institute for Particle Physics\\1156 High St., Santa Cruz, CA 95064, USA}
\affiliation[c]{Department of Astronomy, Cornell University\\Ithaca, NY 14850, USA}

\emailAdd{benvlehmann@gmail.com}
\emailAdd{arwebber@ucsc.edu}
\emailAdd{ogr8@cornell.edu}
\emailAdd{profumo@ucsc.edu}

\abstract{\ignorespaces
   The vast datasets associated with extrasolar systems promise to offer sensitive probes of new physics in the near future. We consider the possibility that such systems may capture primordial black holes (PBHs) or other exotic compact objects, giving rise to unique observational signatures. We estimate the rate of captures by extrasolar systems, accounting for several distinct mechanisms. We find that the capture rate is negligible unless PBHs account for the entirety of dark matter in a narrow mass range just above the threshold of existing constraints from evaporation. In this scenario, luminous evaporating PBHs may be detectable by exoplanet searches.
}


\begin{document}
\maketitle
\flushbottom

\section{Introduction}
\label{sec:introduction}
The future of cosmology and particle physics rests heavily on new astrophysical probes. A growing cast of observational programs offers numerous opportunities to test new physics, even with tools that were designed for entirely different purposes. In particular, new instruments and observational methods have led to surging interest in extrasolar planetary systems within the astronomy community \cite{2007ARA&A..45..397U,Schneider:2011vr,Wright:2010bc,Cassan:2012jx,Akeson:2013mqa,2018ApJS..235...38T}, with potential implications for beyond--Standard-Model (BSM) particle physics. Several recent proposals demonstrate that exoplanets and other small bodies can sensitively probe BSM scenarios, including e.g. dark matter (DM) interactions \cite{Leane:2020wob} and new long-range forces \cite{Tsai:2021irw}. In this work, we study the prospects for using these systems to detect primordial black holes, i.e., black holes that formed at early times from mechanisms besides stellar collapse.

Primordial black holes (PBHs) have long been studied as a potential signature of BSM particle physics \cite{Carr:2003bj,Khlopov:2008qy,Calmet:2014dea}, and are naturally produced in many classes of BSM models \cite{1967SvA....10..602Z,10.1093/mnras/152.1.75,deLavallaz:2010wp,Shandera:2018xkn}. PBHs have been invoked as candidates for the origin of the baryon-antibaryon asymmetry \cite{Garcia-Bellido:2019vlf,Barrow:1990he,Fujita:2014hha,Hook:2014mla,Boudon:2020qpo,Smyth:2021lkn}, the production of particle DM \cite{Fujita:2014hha, Garcia-Bellido:2019vlf, Smyth:2021lkn, Morrison:2018xla}, the source of high-energy photon and cosmic-ray positron emission \cite{Coogan:2020tuf,DeRocco:2019fjq}, and the constituent \emph{per se} of cosmological DM \citep{Carr:2009jm,Carr:2016drx,Lehmann:2018ejc,Carr:2020gox,Green:2020jor,Carr:2020xqk}. If detected, these objects would provide a wealth of information about cosmology and new physics at extremely high scales. Following the observation of binary black hole mergers, there has been renewed interest in PBHs as a DM candidate \cite{Bird:2016dcv,Sasaki:2016jop}, and a wide variety of observational probes have been developed to constrain their abundance across an enormous range of masses. However, despite these constraints, PBHs may yet account for a significant fraction of DM across a broad segment of this mass range \cite{Lehmann:2018ejc}, and in some windows, they can still constitute the entirety of DM \cite{Lehmann:2019zgt,Niikura:2017zjd,Katz:2018zrn,Sugiyama:2019dgt,Montero-Camacho:2019jte,Smyth:2019whb,Jedamzik:2020ypm}.

Many PBH searches are designed to target rare but distinctive signatures. In particular, a \emph{single object} can be identified as a PBH if it lies outside the mass range achievable by stellar collapse, providing clear evidence of new physics and defining a clear direction for subsequent DM searches. Amid this context, it is critical to understand the various astrophysical environments in which one might expect to find PBHs.

If PBHs do make up a significant fraction of cosmological DM, then they should be scattered throughout our galactic DM halo, with a comparable phase space distribution. For particle DM, the phase space distribution is generally sufficient to determine any observable at any time. However, for PBHs, many observables of interest are discrete events that are rare on the timescale of observations. For instance, lensing events \cite{Allsman:2000kg,Wyrzykowski:2011tr,Niikura:2017zjd,Niikura:2019kqi} or low-mass PBH mergers \cite{Abbott:2005pf,Magee:2018opb,Shandera:2018xkn,Authors:2019qbw} would each occur infrequently during the corresponding observations. While the time-averaged rate of such events is determined by the DM phase space distribution, the events themselves are stochastic.

In this work, we consider a scenario which translates this stochasticity from the \emph{timing} of rare events to the distribution of systems in which they occur: we consider the capture of PBHs into bound orbits in an extrasolar stellar system. If PBHs are indeed captured in this manner, and if such captured orbits are long-lived, then some fraction of stellar systems should stably host PBHs at any given time. In the limit that captured objects are permanently bound, a stellar system would only need one encounter with a PBH over its entire history in order to host such an object today. In particular, this means that the rapidly advancing observational techniques probing the dynamics of such systems may also provide a new probe of the PBH population.

Capture requires the incoming PBH to lose mechanical energy in order to become bound to a stellar system. There are several different physical mechanisms that can lead to such a loss of energy, and these mechanisms can be classified by the sink that absorbs energy from the PBH:
\begin{enumerate}
    \item In an encounter with a single other body, the PBH can be rapidly accelerated, causing it to lose energy to gravitational radiation.
    \item In an encounter with a few-body system, the PBH can lose mechanical energy to one object and become bound to another object.
    \item In passing through a many-body system, the PBH can dissipate energy and effectively heat the system.
\end{enumerate}
These energy sinks are each associated with unique phenomenology. In the first case, gravitational waves from such close encounters are potentially detectable directly \cite{Genolini:2020ejw}. In the second case, if the energy transfer is purely mechanical, then the process is time-reversible, and the PBH may be ejected once captured. Finally, in the case of a dissipative process, the deposited energy itself may have observable consequences for the host system.

Making precise predictions for the population of captured objects is inherently challenging. In the case of few-body encounters, such processes are governed by a relatively simple set of parameters, but can nonetheless exhibit complicated chaotic dynamics. By contrast, the dynamics of many-body processes are comparatively simple, but the parameters of these systems are subject to significant astrophysical uncertainties. In this work, we establish order-of-magnitude predictions for the abundance of PBHs captured by each of the above mechanisms across a wide variety of systems.

Throughout this work, we will assume that PBHs make up all of the DM, and we will assume that their velocities are described by a truncated Maxwell--Boltzmann distribution. We focus mainly on two classes of objects: first, black holes at masses near Earth mass $M_\oplus$, which may account for excess microlensing observations by OGLE \cite{Niikura:2019kqi}, and second, microscopic black holes from \SIrange{e-16}{e-12}{M_\odot}, where current constraints are ineffective. In particular, at the lower end of this mass range, active evaporation may make it possible to detect these objects.

This work is organized as follows. We devote one section to each sink of energy that can lead to captures: in \cref{sec:gravitational-waves}, we study the capture of PBHs due to gravitational radiation; in \cref{sec:three-body}, we evaluate the abundance of PBHs captured by few-body interactions; and in \cref{sec:dissipative}, we study captures that take place via dissipative dynamics. We discuss our findings and implications for observables in \cref{sec:discussion}.

\section{Gravitational radiation in two-body encounters}
\label{sec:gravitational-waves}
In this section, we first review basic principles that apply to all captures, and we then estimate the rate and lifetime of captures due to gravitational wave (GW) emission.

\subsection{Generalities of capture}
A capture takes place when a free PBH becomes bound to some stellar system, i.e., when its total mechanical energy changes sign from positive to negative. This requires that the object give up energy to the surroundings. Thus, to evaluate the rate of captures, we can evaluate the cross section for an incoming object to lose an amount of energy commensurate with its initial mechanical energy. For an object that originates far from the stellar system, as a free object should, the initial mechanical energy is simply the initial kinetic energy. Thus, given an initial velocity $v_\infty$, the capture cross section for a PBH of mass $M_\pbh$ is just the cross section for the object to lose an amount of energy greater than its initial kinetic energy:
\begin{equation}
    \label{eq:general-sigma}
    \scap(v_\infty) =
        \int_{\frac12M_\pbh v_\infty^2}^\infty
        \du \loss\frac{\du\scap}{\du \loss}
    .
\end{equation}
We will assume that PBHs have a velocity distribution like that of halo DM, with probability density function (pdf) given by
\begin{equation}
    \label{eq:velocity-distribution}
    f_\infty(v) \propto v^2\exp\left(-\frac{v^2}{v_0^2}\right)
    \Theta\left(\vesc - v\right)
    .
\end{equation}
We take $v_0 = \SI{220}{\kilo\meter/\second}$ and $\vesc = \SI{550}{\kilo\meter/\second}$. This pdf is an approximate description of the equilibrium distribution of DM particles---in our case, PBHs---throughout the halo. Near a point mass like a star, the velocity distribution is modified by the local gravitational potential. Thus, $f_\infty(v)$ should be treated as the distribution of particles far from the stellar system. In particular, the existence of a low-velocity tail of the distribution does not imply that low-velocity objects are ``born'' captured. However, per \cref{eq:general-sigma}, the capture cross section is typically largest for the smallest values of $v_\infty$, and in some cases, the low-velocity tail dominates the capture rate.

Each energy loss mechanism leads to some differential cross section $\du\scap/\du\loss$, and thus to some total cross section $\scap(v_\infty)$. Once these quantities are calculated, the total capture rate is
\begin{equation}
    \Rcap = n_\infty\langle \scap v_\infty\rangle
        = n_\infty\int\du v\,f_\infty(v)\scap(v)v
    ,
\end{equation}
where $n_\infty$ is the number density of objects far from the system and angle brackets denote the average over velocities.
Some systems can also lose captured objects, particularly by ejection in few-body systems with conservative dynamics. In this case, the rate for a particular object to be ejected is independent of the number of objects captured in the system, so we represent this rate by a single quantity $\Rej$. Thus, in a system with $N$ objects captured, the rate for any one object to be lost is $N\Rej$. On a sufficiently long timescale, capture and loss are in equilibrium, meaning that the expected number of captured objects in a system is $\langle N\rangle = \Rcap/\Rej$. Assuming that $\Rej \ll \Rcap$, equilibrium is attained on a timescale of order $t_{\mathrm{eq}} \simeq \langle N\rangle / \Rcap = 1/\Rej$.

\subsection{Capture cross section from gravitational radiation}
We now consider gravitational wave emission as the physical mechanism for energy loss, and evaluate the expected number of objects that are captured by this route.

A PBH that undergoes a close encounter is rapidly accelerated, losing a significant amount of energy to gravitational radiation in the process. If enough energy is lost, the PBH can become bound as a result of the encounter. Such a capture can be much more stable than a capture produced by few-body dynamics. In particular, this process can take place in a \emph{two}-body encounter, or if in a few-body system, it can take place far from the orbital trajectory of the any third body, minimizing the rate of subsequent close encounters that could lead to ejection.

To compute the energy lost to gravitational waves, we follow \refcite{Genolini:2020ejw}. We consider a close encounter between the PBH and a stellar or planetary body $S$. The energy loss in the encounter is given by
\begin{equation}
    \Delta E_{\mathrm{GW}} =
    \frac{8}{15}\frac{M_\pbh^2M_S^2}{(M_\pbh+M_S)^3}
    \frac{p(e)}{(e-1)^{7/2}}
    \times v_\infty^7
    ,
\end{equation}
where $v_\infty$ is the relative speed at infinity, $e$ is the eccentricity of the inbound orbital trajectory, and $p(e)$ is given by
\begin{equation}
    \label{eq:gw-p-e}
    p(e) = (e+1)^{-7/2}\left[
        \arccos\left(-\frac1e\right)\left(
            24 + 73e^2 + \frac{37}{4}e^4 + \frac{\sqrt{e^2 - 1}}{12}\left(
                602 + 673e^2
            \right)
        \right)
    \right]
    .
\end{equation}
The eccentricity can be written in terms of $v_\infty$ and the impact parameter $b$ as
\begin{equation}
    e = \sqrt{1+\frac{b^2v_\infty^4}{(GM_S)^2}}
    .
\end{equation}

Here we are interested in cases in which the PBH is captured, i.e., in which $\Delta E_{\mathrm{GW}}$ exceeds the kinetic energy of the PBH at infinity. We are especially interested in the possibility that the PBH is captured without passing through object $\obj$, so that it does not become captured \emph{within} object $\obj$ and settle to the center. Here there is a very strong dependence on the impact parameter of the encounter, and thus on the radius of object $\obj$. For example, if object $\obj$ is a Jupiter-like planet, then the energy loss will be extremely small for all impact parameters that avoid collisions with the planet. On the other hand, if object $\obj$ is a compact object like a neutron star, then small impact parameters without collisions are indeed realizable, as discussed in detail by \refcite{Genolini:2020ejw}. Such captures are stable on fairly long timescales. In particular, for PBH masses $M_\pbh\lesssim\SI{e-14}{M_\odot}$, these captures survive longer than a Hubble time.

\begin{figure}\centering
    \includegraphics[width=\textwidth]{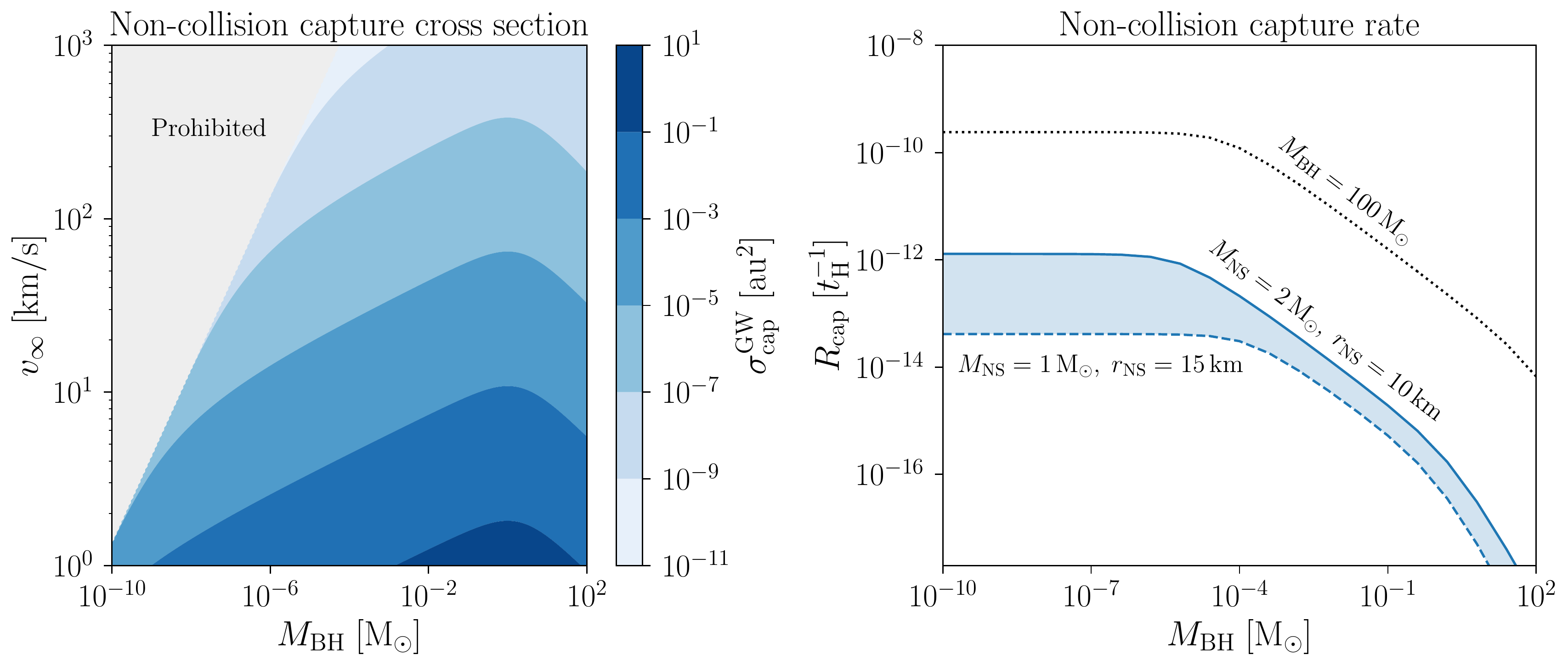}
    \caption{\textit{Left:} Cross section for capture by gravitational wave emission without collision. A neutron star of mass $M_{\mathrm{NS}} = \SI{2}{M_\odot}$ and radius $R_{\mathrm{NS}} = \SI{10}{\kilo\meter}$ is assumed. In the gray region, capture without collision is not possible. \textit{Right:} integrated capture rate as a function of PBH mass in units of the Hubble rate. The shaded area shows the region between two neutron star configurations: one with mass $M_{\mathrm{NS}} = \SI{2}{M_\odot}$ and radius $R_{\mathrm{NS}} = \SI{10}{\kilo\meter}$, and one with mass $M_{\mathrm{NS}} = \SI{1}{M_\odot}$ and radius $R_{\mathrm{NS}} = \SI{15}{\kilo\meter}$. These correspond roughly to the most and least compact neutron stars expected to form \cite{Ozel:2016oaf}. The dotted black curve corresponds to encounters with a \SI{100}{M_\odot} black hole. Substantially less compact objects such as main-sequence stars cannot capture BHs of any size at realistic velocities by GW emission.}
    \label{fig:sigma-gw}
\end{figure}

Captures without collision are possible for a bounded range of impact parameters. The capture condition $\Delta E_{\mathrm{GW}} > \frac12M_\pbh v_\infty^2$ gives a critical impact parameter, $b_{\mathrm{max}}$, below which an encounter will lead to capture. While solving for $b_{\mathrm{max}}$ is in general quite complicated, it is a good approximation to set $e=1$ in \cref{eq:gw-p-e}, corresponding to a free object with minimal kinetic energy. This gives
\begin{equation}
    e = 1 + \left(\frac{bv_\infty^2}{GM_S}\right)^2
        + \mathcal O\left[\left(\frac{bv_\infty^2}{GM_S}\right)^4\right]
    \simeq
    1 +
    \num{e-9}\left(\frac{b}{\SI{100}{\kilo\meter}}\right)^2
    \left(\frac{v_\infty}{\SI{220}{\kilo\meter/\second}}\right)^4
    \left(\frac{M_S}{\SI{1}{M_\odot}}\right)^{-2}
    .
\end{equation}
Taking $e=1$ gives a 7th-order polynomial equation in $b_{\mathrm{max}}$, which is readily solved semi-numerically to find the maximal impact parameter for capture. On the other hand, to avoid a collision, there is a minimum impact parameter: the point of closest approach in a Kepler orbit, $r_{\mathrm{min}}$, is related to the impact parameter by
\begin{equation}
    \label{eq:b-r}
    r_{\mathrm{min}} = \frac{GM_S}{v_\infty^2}\left[
        \left(
            1 + \frac{b^2v_\infty^4}{(GM_S)^2}
        \right)^{1/2}
        -1
    \right]
    ,
\end{equation}
so the minimum impact parameter to avoid a collision is found by setting $r_{\mathrm{min}} = R_S$ in \cref{eq:b-r}, where $R_S$ is the radius of the object $S$. That is, we take
\begin{equation}
    \label{eq:b-min-collision}
    b_{\mathrm{min}} = \sqrt{R_S^2 +\frac{2GM_SR_S}{v_\infty^2}}
    .
\end{equation}
The cross section for capture by gravitational wave emission without collision is then given by $\scap^{\mathrm{GW}} = \pi(b_{\mathrm{max}} - b_{\mathrm{min}})^2\Theta(b_{\mathrm{max}} - b_{\mathrm{min}})$. This cross section is shown as a function of $M_\pbh$ and $v_\infty$ in \cref{fig:sigma-gw}. As expected, $\scap^{\mathrm{GW}}$ is larger for small $v_\infty$, but it also increases moderately for larger PBH massses $M_\pbh$ due to the non-linear dependence on $M_\pbh$ in the energy emitted in GWs.

For our cases of interest, i.e., microscopic PBHs and Earth-mass PBHs, capture by GW emission is exceedingly rare. As shown in \cref{fig:sigma-gw}, the capture rate is extremely small both at small PBH mass, due to the inefficiency of GW emission, and at $M_\pbh \sim M_\oplus$, due to the very small number density of such objects. Given the sharp dependence on the lowest velocities, it is possible that a cold feature in the phase space distribution of the halo could substantially enhance the capture rate, but typical capture rates are well below the Hubble rate (inverse Hubble time). Indeed, the capture rate is below the rate at which captured objects sink to the center of the NS by further GW emission.

\section{Few-body capture and ejection}
\label{sec:three-body}
GW emission is the only mechanism which would allow capture in an encounter between two point masses. Having studied this case, and having demonstrated the generalities of captures, we now move on to mechanisms involving more than two bodies. The simplest possibility is that the PBH loses energy via few-body scattering. In this section, we consider the rate at which PBHs are captured by three-body interactions with extrasolar binary systems.

\subsection{Rates of three-body capture and ejection}
\label{sec:three-body-rates}

Before we evaluate the capture rate, we note that this rate alone does not determine the abundance of bound objects. A major difference between many-body and few-body interactions is that the latter are time-reversible, and thus objects captured in this manner are eventually ejected from the system by the same physics. Describing the population of bound objects thus requires evaluation of both the capture rate and the ejection rate. The ejection rate is difficult to compute directly, since it depends on the complicated physics of the gravitational three-body problem. Moreover, since we take the free PBH population to have a Boltzmann distribution typical of DM, it is necessary to consider the capture of objects across a wide range of velocities. The rate of capture and ejection by three-body interactions has been studied by many authors, including e.g. \refscite{1986AJ.....92..171T,1999Icar..142..509D,2003AsBio...3..207M,Valtonen:2005,Xu:2008ep,Peter:2009mi,Peter:2009mm,Khriplovich:2009jz,Lages:2012hn,Rollin:2014dpa,Lehmann:2020yxb}, and some of these works are readily applicable to the capture of PBHs by generic extrasolar systems.

Here, we use the estimates of the capture and ejection rates provided by \refcite{Lehmann:2020yxb}. In that work, the cross sections for capture and ejection are estimated under a set of approximations that allows for a geometric formulation of the problem. The ejection rate is then determined by modeling subsequent close encounters stochastically. While approximate, the results of \refcite{Lehmann:2020yxb} reproduce the results of numerical simulations while retaining a fully analytical form, allowing them to be used for rapid exploration of a wide variety of systems. Moreover, this formulation accounts for the full dependence of the capture and ejection rates on the parameters of the binary and the velocity of the incoming PBH.

As in \refcite{Lehmann:2020yxb}, we consider the following scenario: a light PBH traverses a system of two objects $A$ and $B$, with $M_A\gg M_B\gg M_\pbh$. The PBH has a close encounter with object $B$, and in the resulting scattering process, the PBH loses energy to object $B$ and becomes bound to object $A$. The resulting bound orbit of the PBH crosses that of object $B$, so they will have additional close encounters. Eventually, such a close encounter transfers enough energy from object $B$ to the PBH that the latter is ejected from the triple system. For simplicity, we take the orbits of objects $A$ and $B$ to be circular, with semimajor axis $R_{AB}$.

The velocity of the PBH just prior to the initial close encounter is denoted by $\bb v_1$. We define the close encounter to begin when the tidal acceleration of the PBH relative to object $B$ due to object $A$ is smaller than its acceleration due to object $B$ by a factor $\epsilon$. As such, $v_1 = |\bb v_1|$ is fully determined by $\epsilon$ and $v_\infty$, the speed of the PBH far from the binary. The direction of the PBH at the close encounter is specified by two angles: the inclination $\beta_1$ of the trajectory relative to the plane of the $AB$ system and the orbital phase $\lambda_1$ of the $AB$ system. The latter is defined with respect to the projection of $\bb v_1$ into the plane of the $AB$ system. For example, when $(\beta_1, \lambda_1) = (0, 0)$, the close encounter takes place with $\bb v_1$ orthogonal to $\bb v_B$, whereas for $(\beta_1, \lambda_1) = (0, \pi/2)$, they are parallel.

Including the effects of gravitational focusing, the cross section for capture of a light PBH by a binary system is given by 
\begin{equation}
    \label{eq:sigma-capture-explicit}
    \scap(\bb v_1) \simeq \frac{v_1}{v_\infty}\frac{\pi (GM_B)^2\left[
        2\vesc^2\left(v_1^{\prime2}+v_B^2\right) - 
        \left(v_1^{\prime2}-v_B^2\right)^2 - \vesc^4
    \right]}{
        \left(v_1^2-\vesc^2\right)^2v_1^{\prime4}
    }
    .
\end{equation}
where $\vesc=\sqrt{2GM_A/R_{AB}}$ is the escape velocity in the potential of object $A$ at the orbital radius of object $B$. This cross section holds only for fixed values of $\beta_1$ and $\lambda_1$, and we must average over these parameters to obtain the cross section averaged over incoming directions. This average is non-trivial because some incoming directions are kinematically prohibited, in which case \cref{eq:sigma-capture-explicit} yields unphysical negative values. Nonetheless, including such unphysical incoming angles leads to a surprisingly robust estimate of capture and ejection rates, and this approach has the advantage that the average can then be written in closed form. Carrying out the average in this simplified fashion yields
\begin{equation}
    \label{eq:sigma-simple-average}
    \widetilde\scap(\bb v_1)\equiv
        \pi\left(\frac{GM_B}{v_1^2-\vesc^2}\right)^2\Biggl[
            -1 - \left(
                \frac{\vesc^2 - v_B^2}{v_1^2 - v_B^2}
            \right)^2 +
            \frac{\vesc^2 + v_B^2}{v_1v_B}\arctanh\left(
                \frac{2v_1v_B}{v_1^2 + v_B^2}
            \right)
        \Biggr]
    .
\end{equation}
\Refcite{Lehmann:2020yxb} extends this approach to derive distributions for the semimajor axis $a$ and eccentricity $e$ of the captured object's orbit after the encounter.

The same methodology can be applied to estimate the timescale on which captured objects are ejected from the system. Such ejection events take place when an energy transfer of comparable magnitude takes place in the opposite direction, from object $B$ to the PBH. \Refcite{Lehmann:2020yxb} uses a simplified application of \"Opik--Arnold theory \cite{1965ApJ...141.1536A} to estimate the ejection rate $\widetilde\Rej(v_\infty)$ of captured objects at fixed $v_\infty$. For our purposes, we are interested in the number of objects expected to be found captured by a particular binary at any given time. This equilibrium number is estimated by
\begin{equation}
    \label{eq:equilibrium-population}
    \langle\widetilde N\rangle = n_\infty\int\du v_\infty\,f(v_\infty)
        \frac{\widetilde\scap(v_\infty)v_1(v_\infty)}{\widetilde\Rej(v_\infty)}
    .
\end{equation}
\Cref{eq:equilibrium-population} can be used to determine whether PBHs at a given mass are typically found captured within binary systems of a particular class: such systems should have $\langle\widetilde N\rangle\gtrsim1$.

\subsection{Compact object capture in different systems}
\label{sec:three-body-optimal}

In the previous subsection, we have outlined a simplified calculation of the capture and ejection rates, and in particular, we have arrived at a relatively simple estimate of the equilibrium population of captured objects. We now consider the classes of systems which are most and least effective at capturing and retaining PBHs.

\begin{figure}\centering
    \includegraphics[width=\textwidth]{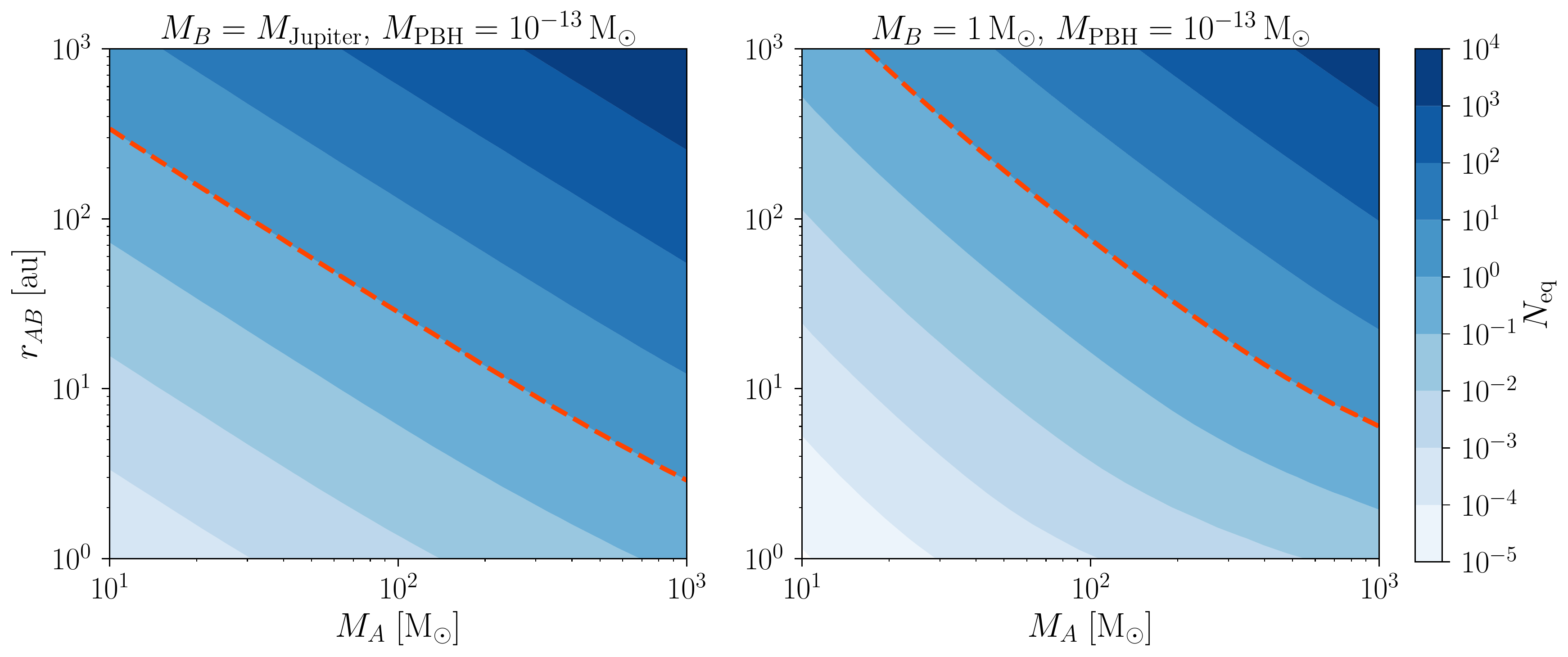}
    \caption{Equilibrium number of captured PBHs as estimated in \cref{eq:equilibrium-population}, assuming all DM is in the form of \SI{e-13}{M_\odot} PBHs, shown as a function of the mass of the heavier object in the binary ($M_A$) and the binary separation ($R_{AB}$). Note that the PBH number density is inversely proportional to $M_\pbh$, so both panels can be recast to other PBH masses by multiplication by $\SI{e-13}{M_\odot}/M_\pbh$. The left and right panels fix $M_B=M_{\mathrm{Jupiter}}$ and $M_B=\SI{1}{M_\odot}$, respectively. In each panel, the dashed red curve indicates $N_{\mathrm{eq}} = 1$. Note that $N_{\mathrm{eq}} \propto M_\pbh^{-1}$ in this regime. The nearly power-law structure of the equilibrium number and the weak dependence on $M_B$ can both be extracted from analytical arguments (see text).
    }
    \label{fig:equilibrium-number}
\end{figure}

The equilibrium number of captured objects is shown in \cref{fig:equilibrium-number} as a function of $M_A$ and $R_{AB}$, for two fixed values of $M_B$. In each panel, it is assumed that all of the DM is in the form of PBHs with mass \SI{e-13}{M_\odot}. As long as the PBH mass is well below the masses of the objects in the binary, the equilibrium number captured scales with their ambient number density, and thus, fixing the DM density, this means that $N_{\mathrm{eq}} \propto M_{\pbh}^{-1}$. The equilibrium number captured increases nearly as a power law with the orbital separation of the binary and with the mass of the heavier object in the system, but is only weakly dependent on the mass of the lighter object. We will explain this behavior shortly. For the moment, we note that with all of DM in the form of \SI{e-13}{M_\odot} PBHs, massive wide binary systems ($M_A\gtrsim\SI{e2}{M_\odot}$, $R_{AB}\gtrsim\SI{100}{\au}$) would \emph{typically} host a large number of captured objects.

To understand the features of \cref{fig:equilibrium-number}, we consider a simpler version of the capture cross section. The capture cross section of \cref{eq:sigma-simple-average} reflects the average over incoming directions. At the order-of-magnitude level, it is analytically simpler to choose a particular orbital phase and inclination angle and evaluate the capture cross section for varying binary parameters. We choose the single configuration that maximizes the energy loss, and thus the capture cross section, for high-velocity compact objects. To identify this configuration, consider the kinematics of three-body captures. When the PBH has a close encounter with object $B$, they can be treated as a two-body system, and in particular the speed of recession of the PBH is equal to the speed of approach in the frame of object $B$. Thus, the maximum energy loss takes place when the direction of the PBH is reversed by the encounter in the frame of object $A$. This is only possible when the PBH velocity before the encounter, $\bb v_1$, is parallel to $\bb v_B$, and the velocity after the encounter, $\bb v_2$, is antiparallel to $\bb v_B$. Taking this directional configuration for $\bb v_1$ and $\bb v_B$ corresponds to fixing $(\lambda_1,\beta_1) = (\pi/2, 0)$. This provides a useful reference point for comparison between different systems. The resulting capture cross section takes a relatively simple form:
\begin{equation}
    \label{eq:sigma-fixed-params}
    \scap(v_\infty) =
    \pi\left(\frac{M_B}{M_A}\right)^2R_{AB}^2
    \times
    \frac{
        1
        + 2\zeta^2\left[
            1 - \left(2 + 2/\zeta^2\right)^{1/2}
        \right]
    }{
        \left[
            1 - \left(2 + 2/\zeta^2\right)^{1/2}
        \right]^4
    }
    \;,
    \qquad
    \zeta = \frac{\vesc}{v_\infty}.
\end{equation}

This expression can be simplified further in the regime relevant for captures: $\scap$ is maximized at $\zeta\approx0.37$, and drops sharply for higher values of $\zeta$, so the capture rate is dominated by objects with $\zeta\ll1$. In this limit, the cross section simplifies to $\scap\simeq \frac\pi4(M_B/M_A)^2R_{AB}^2\zeta^4$. Further, the DM velocity distribution in \cref{eq:equilibrium-population} can be considerably simplified for realistic systems. The capture rate is dominated by the peak in the cross section at $\zeta \approx 0.37$, corresponding to $\mathcal O(1)$ values of $v_\infty / \vesc$. In turn, typical values of $\vesc$ are on the order of \SI{10}{\kilo\meter/\second}, far below $v_0 \approx \SI{220}{\kilo\meter/\second}$. Thus, captures should be dominated by the low-velocity tail of the PBH velocity distribution, which has the form
\begin{equation}
    \label{eq:boltzmann-tail}
    f(v_\infty) \simeq \frac{4v_\infty^2}{\sqrt\pi v_0^3}
    \qquad
    (v_\infty\ll v_0)
    .
\end{equation}
Finally, we fix the semimajor axis and eccentricity of the captured PBH's orbit to representative values $a=3R_{AB}$ and $e=1-R_{AB}/(2a)=5/6$. Together with \cref{eq:equilibrium-population,eq:boltzmann-tail}, this enables a rapid estimate of the equilibrium population of captured objects in a wide variety of systems. Taking $M_B\ll M_A$, the result is
\begin{align}
    \label{eq:velocity-averaged-equilibrium}
    N_{\mathrm{eq}} \simeq
    0.9
    \left(\frac{0.65+\log_{10}\frac{M_A}{M_B}}{3.7}\right)^{-1}
    \left(\frac{M_A}{\SI{1}{M_\odot}}\;\frac{R_{AB}}{\SI{5}{\au}}\right)^{3/2}
    \left(\frac{v_0}{\SI{220}{\kilo\meter/\second}}\right)^{-3}
    \left(\frac{n_\infty\times\SI{e-16}{M_\odot}}
               {\SI{0.3}{\giga\electronvolt/\centi\meter^3}}\right)
    ,
\end{align}
where the base values for the parameters are chosen to be roughly representative of the capture of objects of mass \SI{e-16}{M_\odot} by the Sun--Jupiter system, assuming they account for the entirety of the local DM density. Strictly speaking, this is an estimate of an upper bound on the capture rate, since the angles $\lambda_1$ and $\beta_1$ are chosen in the most favorable configuration possible. Nonetheless, this serves as an informative estimate of the capture rate at the order of magnitude level and applies to a wide range of systems. Indeed, this result is a reasonably good match to the numerical results in \cref{fig:equilibrium-number}, overestimating the number of captured objects by about an order of magnitude.

PBHs at masses below $\sim$\SI{e-16}{M_\odot} are strongly constrained by evaporation, so this optimistic estimate indicates that capture in the Sun--Jupiter system is only possible for PBH DM in a narrow mass range. Nonetheless, this estimate suggests that if a substantial fraction of the DM is composed of PBHs with significant evaporation luminosity, then it is possible that a bright point source could be found captured within the solar system. Recently, \refcite{Baker:2021btk} studied the potential implications of discovering such a low-mass PBH nearby: since such an object would be actively evaporating, the relationship between the object's mass and evaporation rate would enable a direct count of the number of dark-sector degrees of freedom. Our calculation suggests that if a population of such objects were maintained for a sufficiently long time, then there would be good prospects to find one close enough to be studied in this manner. However, since such a measurement relies on the rapid evaporation of the observed PBH, such a population would not be stable on cosmological timescales.

In the limit of asymmetric masses $M_B \ll M_A$, the equilibrium number of captured objects is only very weakly dependent on the mass of the lighter object in the binary. This is to be expected: in such a case, the cross sections for capture and ejection both scale with $M_B^2$. On the other hand, systems with larger $M_A$ and $R_{AB}$ are much more efficient at capturing and retaining light PBHs. At the upper ranges of \cref{fig:equilibrium-number}, a wide binary with a \SI{100}{M_\odot} central object and an orbital separation of \SI{e3}{\au} has $\langle\widetilde N\rangle \sim \num{e3}$ for all of DM in the form of PBHs with mass \SI{e-13}{M_\odot}. Thus, such a system has an $\mathcal O(1)$ probability of hosting at least one PBH in a bound orbit if the PBH mass is below \SI{e-10}{M_\odot}. The capture rate of Earth-mass objects is very low in all realistic binary systems, so such captured objects cannot account for OGLE microlensing events. If instead DM is in the form of light PBHs with mass between \SI{e-16}{M_\odot} and \SI{e-14}{M_\odot}, as is allowed by current constraints, then such objects should be commonly bound in systems only slightly heavier and wider than the Solar system.

\section{Dissipative dynamics}
\label{sec:dissipative}
The treatment of the previous section is limited to capture by three-body interactions. We now turn our attention to capture by many-body interactions, which are qualitatively distinct due to dissipation: such captures are not time-reversible. When PBHs are captured around single objects, ejection is impossible. Even in multi-component systems, dissipative captures are much less prone to ejection than their few-body counterparts. Thus, even though the rates of dissipative captures are naively much smaller, it is important to evaluate their contribution to the population of bound objects.

\subsection{Gas drag and dynamical friction}
As an unbound object such as a planet passes through a gaseous environment, its kinetic energy is dissipated via interactions with many gas particles, potentially resulting in a capture \cite{1979Icar...37..587P}. A similar mechanism may lead to captures of certain types of compact objects. However, only a particular class of compact objects are subject to the usual physics of gas drag. As usually treated, gas drag presumes that the object efficiently displaces gas in its path, but this is not the case for dark compact objects such as PBHs.

A black hole will still accrete gas particles in its path, which will reduce the object's specific kinetic energy. However, this effect is suppressed by the very small geometric cross section of the black hole. Including gravitational focusing, this cross section is
\begin{equation}
    \sigma_\pbh =
        \pi\left(\frac{2GM_\pbh}{c^2}\right)^2
        \Bigl(1 + \frac{c^2}{v_{\mathrm{rel}}^2}\Bigr)
    ,
\end{equation}
where we have used the Schwarzschild radius $r_\pbh = 2GM_\pbh/c^2$. Now suppose that a black hole with initial velocity $v_\infty$ transits through a spherical gas cloud of density $\rho$ and radius $R$, accreting a mass $\Delta M_\pbh \approx 2\rho R\sigma_\pbh$, and suppose that the accreted gas particles are slow compared to the accreting PBH. At the end of the transit, the potential energy is reduced by $\Delta U_\pbh \simeq -GM_{\mathrm{cloud}}\,\Delta M/R$ due to the accreted mass. Capture requires that the total mechanical energy becomes negative, and since the accreted mass leaves the kinetic energy constant, the potential energy must decrease by at least $T_\infty = \frac12M_\pbh v_\infty^2$. Taking $v_{\mathrm{rel}} \ll c$, and neglecting changes in the PBH velocity due to accretion, we have
\begin{equation}
    \frac{\Delta U_\pbh}{T_\infty} \simeq
    \frac{32\pi^{3/2}G^3\rho^2R^2M_\pbh}{v_\infty^2c^2\left[
        3G\rho\left(v_\infty^2+4\pi G\rho R^2\right)
    \right]^{1/2}}\arctanh\left[
        2R\left(\frac{\pi G\rho}{3(v_\infty^2+4\pi G\rho R^2)}\right)^{1/2}
    \right]
    .
\end{equation}
For massive clouds with $4\pi G\rho R^2 \gg v_\infty^2$, this simplifies to
\begin{equation}
    \frac{\Delta U_\pbh}{T_\infty} \simeq
    \num{e-13}\times
    \left(\frac{M_\pbh}{M_\oplus}\right)
    \left(\frac{R}{\SI{e5}{\au}}\right)
    \left(\frac{\rho}{\SI{e-10}{\gram/\centi\meter^3}}\right)
    \left(\frac{v_\infty}{\SI{220}{\kilo\meter/\second}}\right)^{-2}
    .
\end{equation}
Thus, even for densities and radii well in excess of those of typical gas clouds, simple accretion is not an efficient energy loss mechanism for black holes.

However, black holes are still subject to dynamical friction, i.e., energy loss due to gravitational interactions with the gas, and we now estimate the rate of captures by this mechanism. First, consider a perturber of mass $M_\pbh$ moving with velocity $v$ in the rest frame of a uniform gaseous medium with density $\rho$. Dynamical friction in this scenario has beeen treated by \refcite{Ostriker:1998fa}, and previously applied to planet formation by \refcite{2015ApJ...811...54G}. The frictional force on the perturber depends first on whether the relative velocity is subsonic or supersonic. Recall that for an ideal gas with adiabatic index $\gamma$ and molecular mass $m_{\mathrm{mol}}$, the sound speed is given by $c_s^2 = \gamma\kB T/m_{\mathrm{mol}}$. In terms of the Mach number $\mach\equiv v/c_s$, the frictional force is given by
\begin{equation}
    F_{\mathrm{DF}} = - \frac{2\pi\rho G^2M_\pbh^2}{\mach^2c_s^2}
    \begin{cases}
        \displaystyle
        \log\left(\frac{1+\mach}{1-\mach}\right) - \mach
            & \mach < 1 \\
        \displaystyle
        \log\left(1-1/\mach^{2}\right) + 
            2\log\left(\frac{d_{\mathrm{max}}}{d_{\mathrm{min}}}\right)
            & \mach > 1.
    \end{cases}
\end{equation}
Here $d_{\mathrm{min}}$ is the distance of closest approach between gas molecules and the perturber, and $d_{\mathrm{max}}$ is the length scale of the wake left behind as the object traverses the medium. For macroscopic objects, $d_{\mathrm{min}}$ is cut off by the size of the perturber itself. In our case, working with compact objects, the size of the perturber can be very small: a black hole of mass \SI{e-9}{M_\odot} has a Schwarzschild radius on the order of \SI{3}{\micro\meter}. Depending on the black hole mass and gas density, the Schwarzschild radius $R_\pbh$ may be smaller than the typical spacing of the gas molecules, $d_{\mathrm{mol}}\equiv\left(\rho/m_{\mathrm{mol}}\right)^{-1/3}\approx \SI{0.5}{\micro\meter}\left[\rho/(\SI{e-8}{\kilo\gram/\meter^3})\right]^{-1/3}\left[m_{\mathrm{mol}}/m_{\mathrm H}\right]^{1/3}$, where $m_{\mathrm H}$ is the mass of Hydrogen. We take $d_{\mathrm{min}}$ to be the larger of these two scales, $d_{\mathrm{min}} = \max\{R_\pbh, d_{\mathrm{mol}}\}$. We set $d_{\mathrm{max}} = vt$ a time $t$ after the perturber enters the cloud, and we neglect times for which $d_{\mathrm{max}} < d_{\mathrm{min}}$.

Note that the dynamical friction force is proportional to $M_\pbh^2$, so the acceleration of the perturber is linear in the perturber's mass. However, if the mass density of PBHs is held fixed, the number density scales as $M_\pbh^{-1}$, so the capture rate should be approximately independent of mass in this case. This independence is not exact due to the weak logarithmic dependence on $M_\pbh$ via $d_{\mathrm{min}}$ in the regime where the latter is set by the Schwarzschild radius.

Now we specialize to a uniform spherical cloud of radius $R$, and assume that the perturber travels through the center of the cloud. The energy lost over the course of the encounter is $\Delta E = \int_{-R}^R\du s\,F_{\mathrm{DF}}(s)$, where $s$ parametrizes the trajectory. Anticipating that $\Delta E/E$ is small, we neglect any change in $\mach$, so the only $r$-dependence in $F_{\mathrm{DF}}$ comes from setting $d_{\mathrm{max}} = vt = s + R$. Then the fractional energy loss is
\begin{equation}
    \label{eq:fractional-loss}
    \frac{\Delta E}{E} = \frac{8\pi^2\rho G^2 M_\pbh R}{\mach^{4}c_s^4}
    \begin{cases}
        \log\left(\frac{1+\mach}{1-\mach}\right) - \mach
            & \mach < 1\\
        \log\left(1 - 1/\mach^2\right)
        + 2\left[
            \log\left(
                \frac{2R}{d_{\mathrm{min}}}
            \right) - 1
        \right] & \mach > 1.
    \end{cases}
\end{equation}
In the far subsonic and supersonic regimes, this reduces to
\begin{equation}
    \frac{\Delta E}{E} \simeq \frac{8\pi^2\rho G^2 M_\pbh R}{c_s^4}\begin{cases}
        \mach^{-3}
            & \mach \ll 1\\
        2\left[
            \log\left(
                \frac{2R}{d_{\mathrm{min}}}
            \right) - 1
        \right]\mach^{-4} & \mach \gg 1.
    \end{cases}
\end{equation}
Note that $\mach$ is bounded below due to acceleration by the cloud itself: an object with $v_\infty > 0$ will enter the cloud with velocity above $\mach_{\mathrm{min}} \equiv \left[8\pi G\rho/3\right]^{1/2}R/c_s$.

A typical nebula has a number density of order \SI{e3}{\centi\meter^{-3}}, an extent of at most \SI{1}{\parsec}, and a temperature $T \sim \SI{e4}{\kelvin}$ \cite{1980A&A....89..336P,1997A&A...318..256G}. The speed of sound is then $c_s \approx \sqrt{(5/3)\kB T/m_{\mathrm{H}}} \sim \SI{10}{\kilo\meter/\second}$ for Hydrogen, while $\mach_{\mathrm{min}} \sim 0.09$ for the same configuration. Since $\mach_{\mathrm{min}} \ll 1$, we first consider the subsonic regime. In the subsonic limit, the capture condition $\Delta E/E > 1$ becomes $\mach < \mach_{\mathrm{cap}} \equiv 2\left(\pi^2\rho R G^2M_\pbh/c_s^4\right)^{1/3}$, and imposing $\mach_{\mathrm{cap}} > \mach_{\mathrm{min}}$ leads to the requirement
\begin{multline}
    1 < \frac{\mach_{\mathrm{cap}}}{\mach_{\mathrm{min}}}
    =
    \left(
        \frac{27\pi}{8}\frac{GM_\pbh^2}{\rho c_s^2R^4}
    \right)^{1/6}
    \\
    \approx
        \num{0.002}
        \left(\frac{M_\pbh}{\SI{1}{M_\oplus}}\right)^{1/3}
        \left(
            \frac{\rho}{\num{e3}m_{\mathrm{H}}/\SI{}{\centi\meter^3}}
        \right)^{-1/6}
        \left(\frac{R}{\SI{1}{\parsec}}\right)^{-2/3}
        \left(\frac{c_s}{\SI{10}{\kilo\meter/\second}}\right)^{-1/3}
    .
\end{multline}
Thus, far-subsonic captures require unrealistically large masses, small system dimensions, or low sound speeds. Note that as $R,\rho\to0$, although $\mach_{\mathrm{cap}} / \mach_{\mathrm{min}}$ becomes large, $\mach_{\mathrm{cap}}$ vanishes, so small or low-density systems can only capture objects in subsonic transits for an extremely narrow range of initial velocities.

In the supersonic limit, $\Delta E/E$ is suppressed by $\mach^4$, so the most promising regime for captures is the transonic regime, $\mach \approx 1$. The dynamical friction force peaks here, with a divergence at $\mach = 1$. The force in the transonic regime is well approximated by taking $F_{\mathrm{DF}}$ to be symmetric about $\mach=1$ and expanding the $\mach < 1$ expression about that point. This gives
\begin{equation}
    F_{\mathrm{DF}} \simeq - \frac{2\pi \rho G^2M_\pbh^2}{v^2}\left[
        -1 + \log\left(\frac{2}{\left|\mach-1\right|}\right)
    \right]
    ,
\end{equation}
with $\Delta E \simeq 2R F_{\mathrm{DF}}$. In the transonic limit, one can solve for the maximal value of $\left|\mach - 1\right|$ leading to capture, finding
\begin{multline}
    \left|\mach - 1\right| < 2\exp\left(
        -1 - \frac{c_s^4}{8\pi \rho G^2R M_\pbh}
    \right)
    \\
    \approx 2\exp\left[
        \num{-3e11}
        \left(\frac{c_s}{\SI{10}{\kilo\meter/\second}}\right)^4
        \left(\frac{R}{\SI{1}{\parsec}}\right)^{-1}
        \left(\frac{M_\pbh}{\SI{1}{M_\oplus}}\right)^{-1}
        \left(
            \frac{\rho}{\num{e3}m_{\mathrm{H}}/\SI{}{\centi\meter^3}}
        \right)^{-1}
    \right]
    .
\end{multline}
Thus, for any realistic parameter values, transonic captures are viable only for a vanishingly narrow range of initial velocities.

\subsection{Collisions with stellar and planetary bodies}

\begin{figure}\centering
    \makebox[\textwidth][c]{
        \includegraphics[width=1.1\textwidth]{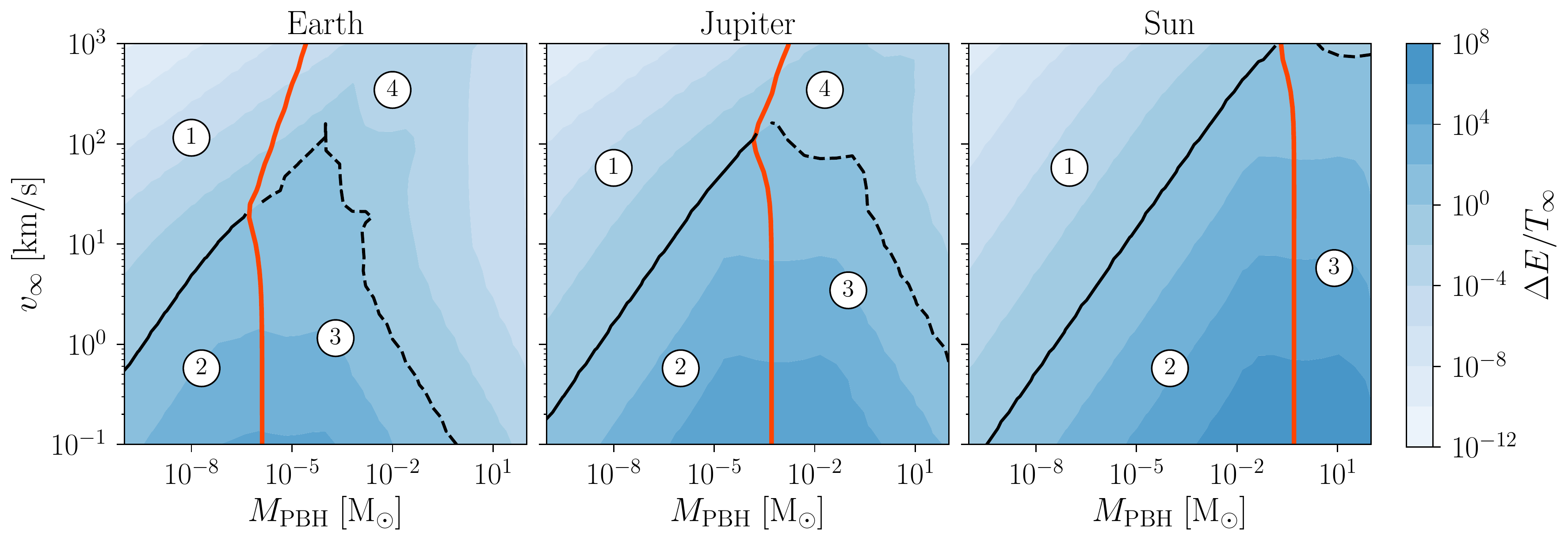}
    }
    \caption{Energy loss in transiting through a body as a fraction of the kinetic energy at infinity, as a function of the PBH mass and velocity at infinity. Panels show three benchmark cases with parameters of Earth, Jupiter, and the Sun. In each panel, points below the black curve result in captures ($\Delta E > T_\infty$), and points to the right of the orange curve destroy the target object ($\Delta E$ exceeds the binding energy). The four labeled points correspond to the following scenarios: \textbf{(1)} the transit neither captures the PBH nor destroys the target; \textbf{(2)} the transit captures the PBH without destroying the target; \textbf{(3)} the target is destroyed, and the PBH is bound to the system; \textbf{(4)} the target is destroyed and the PBH remains free.}
    \label{fig:transits}
\end{figure}

In the previous subsection, we have considered PBH capture by dissipation in a gas cloud, where the geometric cross section is large compared to the three-body close encounters of \cref{sec:three-body}, but the typical energy losses are much smaller. However, thus far, we have neglected a mechanism for large dissipative losses with a small cross section: transit of a PBH through a planet or star. In this scenario, the PBH dissipates energy by the same dynamical friction mechanism that drives losses in a gas cloud, but the higher density of a planet or star leads to much more significant energy losses during such a transit. We now consider the population of objects that would be captured by this particular mechanism.

The energy lost to dynamical friction in passing through a star or planet can be computed by a similar procedure as in the previous section, but our treatment now differs in two ways. First, since the typical parameter values are quite different from those of gas clouds, we do not assume that the PBH velocity is constant throughout the encounter. Instead, we compute the instantaneous energy loss by numerical solution of the equations of motion. Second, since there may be observational implications, we are motivated to consider the destruction of objects by the passage of PBHs in addition to capture. We perform a simplified treatment of planet and star destruction: we say that the target is destroyed if the energy lost by the PBH to dynamical friction, $E_{\mathrm{DF}}$, exceeds the binding energy of the target, $U_G = -3GM_{\target}^2/(5R_{\target})$.

We study energy losses in three benchmark systems: Earth, Jupiter, and the Sun. The results are shown in \cref{fig:transits}. For each of these cases, capture of a PBH without destruction of the target is possible for sufficiently light PBHs with low initial velocities, i.e., in region \textbf{(2)} of each panel. For higher initial velocities, the encounter takes place with $\mach \gg 1$, and the dynamical friction force is suppressed. Similarly, large PBH masses $M_\pbh \gg M_{\target}$ accelerate the target and guarantee a highly supersonic encounter, so the energy loss at large PBH masses is negligible and capture is impossible. Observe that in all three cases, destruction of the target requires $M_\pbh \gtrsim M_{\target}$: upon collision, the kinetic energy of a PBH falling from rest is given by $GM_{\target}M_\pbh/R_{\target}$, which only exceeds $U_G$ if $M_\pbh > (3/5)M_{\target}$. Thus, whenever destruction is possible, the ratio of the PBH number density to the target number density is bounded above by the ratio of their volume-averaged mass densities, $n_\pbh/n_{\target} \lesssim \rho_\pbh / \rho_{\target}$. This factor in turn is $\mathcal O(100)$ for stars, which suggests that stellar destruction events take place on a timescale at most a factor of $\mathcal O(100)$ shorter than that for collisions of stars, which are exceedingly rare. Given that the number density of planets is parametrically close to the number density of stars, the maximum rate of destruction events in these systems at first appears to be higher by a factor $M_{\mathrm{planet}} / M_{\mathrm{star}}$, but the geometric cross section suffers a comparable suppression.

Thus, destruction events of any kind are rare. Explicit computation confirms that the destruction rate is comparable across our three benchmark systems, and is no higher than \SI{e-26}{\second^{-1}} for any PBH mass, well below the Hubble rate. Captures are also extremely rare and do not occur in excess of \SI{e-24}{\second^{-1}}. Captured objects can undergo subsequent transits, which in principle enhances the destruction rate, but not above the very low capture rate. Note that destruction by PBH encounters has been previously considered by \refcite{Abramowicz:2008df} in the context of luminous signatures of PBH collisions with stellar and planetary objects, with qualitatively similar results. Stellar capture of DM has likewise been studied previously e.g. by \refscite{Nussinov:2009ft,Jungman:1994jr}.

\subsection{Adiabatic contraction}
\label{sec:dissipative-adiabatic-contraction}
A third possibility is that dissipation of gravitational energy of the gas itself provides a mechanism for the capture of dark compact objects. As gas collapses during star formation, the potential well deepens, and nearby objects can thus be captured---not by a loss of kinetic energy, but by a reduction in potential energy. A key element of the process is that as the gas density increases, the local DM density is gravitationally enhanced, a process known as adiabatic contraction \cite{Gnedin:2004cx}. Thus, during the process of star formation, DM particles---or equivalently, dark compact objects such as PBHs---can be efficiently captured.

This mechanism has been considered in detail by \refcite{1999A&A...343....1D} for its effects on the population of luminous evaporating black holes captured around stars, and more recently by \refscite{Capela:2012jz,Capela:2013yf,Capela:2014ita} in the context of stellar destruction. Following \refcite{Capela:2012jz}, a gas cloud of density $\rho_{\mathrm{g}}$ and radius $R_{\mathrm{g}}$ captures a DM halo with density of order
\begin{equation}
    \rho_{\mathrm{bound}} \simeq \rho_{\mathrm{DM}}\times\frac{4\pi}{3}\left(
        \frac{6G\rho_{\mathrm{g}} R_{\mathrm{g}}^2}{\frac32v_0^2}
    \right)^{3/2}
    ,
\end{equation}
where $v_0$ is the characteristic DM velocity dispersion of \cref{eq:velocity-distribution}. Due to adiabatic contraction, the bound DM particles (compact objects) assume an equilibrium distribution with a power-law profile $\rho_{\mathrm{bound}}(r)\sim r^{-3/2}$. We assume that the extent of the bound halo is the cloud radius $R$, so that the number density within any particular radius can be readily calculated.

The baryonic gas that forms stars is at first found in giant molecular clouds, with masses as large as \SI{e6}{M_\odot} and radii as large as \SI{10}{\parsec} \cite{Heyer:2015}. These clouds fragment and form many prestellar cores, with typical masses of $1\textnormal{--}\SI{10}{M_\odot}$ and typical radii of $\num{3000}\textnormal{--}\SI{6000}{\au}$ \cite{Kirk:2005ng}. Even for a dense system with a total mass of \SI{10}{M_\odot} and a radius of \SI{3000}{\au}, with $\rho_{\mathrm{DM}} = \SI{0.3}{\giga\electronvolt/\centi\meter^3}$, the density of bound DM is negligible, $\rho_{\mathrm{bound}} \approx \SI{6e-7}{\giga\electronvolt/\centi\meter^3}$. This is mainly due to the sharp dependence on the velocity dispersion $v_0$: if the system under consideration forms in a small DM substructure with a small dispersion, then the bound density can be considerable. In particular, in globular clusters, constraints can be derived from the absence of stellar destruction, as in \refcite{Capela:2012jz}. However, for a generic stellar system, capture due to adiabatic contraction is negligible.

\section{Discussion}
\label{sec:discussion}

In the preceding sections, we have studied several distinct mechanisms by which PBHs can be captured in stellar systems. Some of these mechanisms give rise to bound orbits which are potentially short-lived, ending in ejection from the system or accretion into another body. Others produce stable, long-lived orbits, partially compensating for smaller capture cross sections. We have also evaluated the rate of destruction of planetary or stellar bodies by PBH encounters, and this is guaranteed to be negligible, requiring fairly high PBH masses and thus low number densities.

The most interesting captures are those which give rise to clear observables. More massive PBHs, particularly in the OGLE mass range, $\sim$\SI{e-6}{M_\odot}, would be more easily detected in extrasolar systems. Such PBHs are comparably massive to planets, so any observable must discriminate between such light PBHs and planets of the same mass. It is possible that such objects could be distinguished from planets based on the \textit{absence} of stellar occultations. Occultation events, in which a star is dimmed by the transit of a planet, are a key non-gravitational signature used to detect planets in exoplanet searches. If \textit{gravitational} Doppler shifting is observed to occur with a statistical excess compared to occultations, this would signal the presence of compact objects that do not block light.

Still, this strategy can only be effective in a class of systems where the expected number of captured objects is at least comparable to the number of planets. On the contrary, our results indicate that the capture of such massive PBHs is exceedingly rare. The rate of captures is suppressed by the number density of PBHs, which is very small even in the most optimistic cases. Even if objects with mass as low as \SI{e-8}{M_\odot} could be reliably detected by gravitational means, and even if they accounted for 10\% of the DM density, the equilibrium values in \cref{fig:equilibrium-number} would still be suppressed by at least \num{e-6}. This implies that $\langle N \rangle \ll 1$ even in the widest and most massive binaries, making this an unlikely probe of the PBH population.

Rather, the capture rate is inevitably highest for DM composed of light PBHs, in the open window in the constraints for $\SI{e-16}{M_\odot}\lesssim M_\pbh\lesssim\SI{e-12}{M_\odot}$. We have shown that such objects can be frequently captured in realistic stellar systems, particularly massive wide binaries. However, in most such systems, it is improbable that such a light object would produce a distinctive observable signature: objects with non-negligible capture rates would be comparable in mass to asteroids or even lighter.

For these lighter objects, there are still two observables of interest. First, there are potential implications for pulsar timing signatures. It is known that the timing signature can be observably perturbed by short-lived PBH flybys \cite{Dror:2019twh}. In our scenario, it is also possible to witness a short-lived capture. Here, a PBH has a close encounter with a binary companion of a millisecond pulsar, and is captured into a short-lived bound orbit before being ejected from the triple system. Such captures are almost always short-lived because of the small semimajor axis of pulsar binaries, and the cross section for such captures is extremely small. Indeed, the assumptions of \cref{sec:three-body} are typically violated in such systems, and simulations suggest that the rate is an order of magnitude smaller than the prediction of \cref{eq:sigma-capture-explicit}. Nonetheless, such temporary captures would have a distinctive signature in pulsar timing. In particular, the signatures and population statistics of pulsar planets have been studied previously in the astronomy community (see e.g. \refscite{1992ApJ...387L..69T,2016ApJ...832..122M}). A captured PBH would result in the temporary appearance of a pulsar planet, which would vanish on the timescale of $\mathcal O(\SI{1}{\year})$ once the object is ejected.

Second, at the very lightest end of the allowed mass range, such black holes would be actively evaporating today. Thus, systems which capture PBHs are a promising environment in which to search for PBH evaporation. As extrasolar systems are probed by a new generation of telescopes as part of the rapidly accelerating exoplanet program, it is possible in principle to see evidence of PBH evaporation using the same instruments. PBHs at the lowest masses consistent with present-day constraints would produce radiation at MeV energies and below, with significant emission even down to optical wavelengths (see e.g. \refcite{Coogan:2020tuf}). We conclude that the best prospect for the observation of a captured PBH would be the detection of evaporation signatures by an exoplanet search. However, at present, we can draw no additional constraints on low-mass PBHs.

\acknowledgments
BVL is supported by DOE grant DE-SC0010107 and by the Josephine de Karman Fellowship Trust. AW is supported by an Undergraduate Research in Science and Technology Award and by the Koret Foundation. SP is supported in part by DOE grant DE-SC0010107. BVL thanks the Racah Institute of Physics at the Hebrew University of Jerusalem for hospitality during completion of this work.

\bibliographystyle{JHEP}
\bibliography{references}

\end{document}